# Chalcogenide-glass polarization-maintaining photonic crystal fiber for mid-infrared supercontinuum generation

A. N. GHOSH,[1,*] M. MENEGHETTI,[2] C. R. PETERSEN,[3,5] O. BANG,[3,5] L. BRILLAND,[4] S. VENCK,[4] J. TROLES,[2] J. M. DUDLEY,[1] AND T. SYLVESTRE[1]

[1] *Institut FEMTO-ST, CNRS, Université Bourgogne Franche-Comté UMR6174, Besançon, France*
[2] *Univ Rennes, CNRS, ISCR-UMR 6226, F-35000 Rennes, France*
[3] *DTU Fotonik, Dept. of Photonics Engineering, Technical University of Denmark, Lyngby, Denmark*
[4] *SelenOptics, 263 Avenue du Gal Leclerc, Campus de Beaulieu, 35700 Rennes, France*
[5] *NORBLIS IVS, Virumgade 35D, 2830 Virum, Denmark*
*\*amarnath.ghosh@femto-st.fr*

**Abstract:** In this paper, we report the design and fabrication of a highly birefringent polarization-maintaining photonic crystal fiber (PM-PCF) made from chalcogenide glass, and its application to linearly-polarized supercontinuum (SC) generation in the mid-infrared region. The fiber was drawn using the casting method from $As_{38}Se_{62}$ glass which features a transmission window from 2 to 10 µm and a high nonlinear index of $1.13 \times 10^{-17}$ m$^2$/W. It has a zero-dispersion wavelength around 4.5 µm and, at this wavelength, a large birefringence of $6 \times 10^{-4}$ and consequently strong polarization maintaining properties are expected. Using this fiber, we experimentally demonstrate supercontinuum generation spanning from 3.1-6.02 µm and 3.33-5.78 µm using femtosecond pumping at 4 µm and 4.53 µm, respectively. We further investigate the supercontinuum bandwidth versus the input pump polarization angle and we show very good agreement with numerical simulations of the two-polarization model based on two coupled generalized nonlinear Schrödinger equations.

## 1. Introduction

Supercontinuum (SC) generation towards the mid-infrared (MIR) range is a very active field of research motivated by a wide range of applications including optical coherence tomography (OCT), material processing, optical sensing and absorption spectroscopy [1-14]. Although much progress in this field has been reported recently, state-of-the-art mid-IR SC systems are still in their infancy and considerable effort is required before mature broadband IR SC light sources beyond 5 µm are made available for industrial applications. Soft glasses, such as chalcogenide ($As_2S_3$, $As_2Se_3$, GeAsSe) [1-5], Tellurite ($TeO_2$), chalcohalides (GeTe, GeAsTeSe) [15,16], heavy metal oxide ($PbO$-$Bi_2O_3$-$Ga_2O_3$-$SiO_2$-$CdO$) [7] and ZBLAN ($ZrF_4$–$BaF_2$–$LaF_3$–$AlF_3$–$NaF$) [17-19] have been widely used for drawing highly nonlinear infrared fibers, and experiments have shown efficient SC generation in the mid-IR up to 15 µm [1,2], and up to 11 µm using all-fiber cascaded systems [20,21]. Among the wide variety of infrared fibers, chalcogenide-glass-based optical fibers (composed of S, Se or Te) are excellent photonic platforms for nonlinear applications in the mid-IR due to their wider transmission window, tailorable dispersion in the mid-IR, and high optical nonlinearity up to hundred times greater than silica or ZBLAN glasses [1-4,22].

On the other hand, there is particular interest in combining the chalcogenide fiber platform with polarization-preserving properties as this will enable polarization-dependent applications in interferometric techniques, gas and pressure sensing, integrated-optic devices and optical coherence tomography (OCT) [23-26]. Polarization-maintaining fibers also help in minimizing detrimental effects such as polarization noise and instability [27]. There are several ways of introducing strong birefringence in the fiber for polarization maintaining propagation. For instance, birefringence can be obtained by applying mechanical stress in the cladding thus inducing anisotropy of the refractive index in the core region, which results in a modal birefringence up to $5\times10^{-4}$ for the so-called *PANDA* fibers [28-30] and close to 60% higher birefringence in *bow-tie* fibers [31]. Highly birefringent fiber can also be fabricated using the powder-in-tube process where two rods of glass material are placed on the sides of the core [32]. Another way of obtaining high birefringence is by breaking the symmetry of the fiber structure. This was achieved in photonic crystal fibers (PCFs) where an asymmetric arrangement of air holes is designed with two different diameter air holes located in orthogonal position near the core of the fiber, thus providing higher effective index difference between the two orthogonal polarization modes [33-36]. In addition to the alteration of air hole diameter in the core region of PCFs, an asymmetric structure can also be obtained by either introducing mechanical stress [38] or modifying the shape of air holes [39]. It has been shown that the birefringence of silica based PM-PCFs can reach high values up to $10^{-3}$ [34, 36-37], which is one order of magnitude higher than that of conventional PANDA or bowtie silica fibers [28-31].

Here we report the design and fabrication of a chalcogenide glass-based highly birefringent PM-PCF for SC generation in the mid-infrared. Numerical modelling is carried out in order to compute the birefringence and dispersion properties of the fiber. Two supercontinuum spectra spanning from 3.1-6.02 µm and 3.33-5.78 µm were experimentally demonstrated using a MHz femtosecond optical parametric oscillator with a central wavelength of 4 µm and 4.53 µm, respectively. The polarization dependence of the supercontinuum when pumping at 4.53 µm was also studied, and the experimental spectra were checked against numerical simulations of the two-polarization model based on two coupled generalized nonlinear Schrödinger equations. We furthermore report on the fabrication of long PM-PCF tapers directly on the drawing tower.

The remainder of this paper is organized as follows: in section 2 we outline the methods for chalcogenide glass fabrication, preform, PCF and taper drawing. In section 3 we report numerical simulations of the microstructured optical fibers and provide both the birefringence and the group-velocity dispersion (GVD). We show in particular how we can tailor both the birefringence and the dispersion by tapering the PM PCF down to a few µm core diameter. In section 4 we provide the transmission and attenuation spectra of the PM PCFs. In section 5 we describe the experimental setup for mid-IR SC generation and in section 6 we show experimental results of broadband SC spectra at different pump wavelength. The effect of pump polarization angle is finally investigated in section 7 both experimentally and numerically using two-coupled generalized nonlinear Schrödinger equations (GNLSE).

## 2. Preform and fiber fabrication

### 2.1 Glass fabrication

The fiber preform was made at the University of Rennes from $As_{38}Se_{62}$ chalcogenide glass. It was produced using the conventional melt-quenching method where a silica ampoule is filled with high purity As (99.999%), Se (99.999%) and oxygen and hydrogen getters like $TeCl_4$ and Mg [40-42]. The silica tube is then placed in a two-stage vacuum pump (combination of oil pump and turbo molecular pump) where it is pumped for approximately 3 hours. Next, the silica ampoule is sealed and then placed inside a furnace for 12 hours at 850°C. The furnace has a rotational motion that enables uniform mixture of molten materials. During the heating process, $TeCl_4$ produces HCl and $CCl_4$ by collecting additional hydrogen and carbon, thus increasing the purity of the glass. After that, the ampoule is cooled down to 700°C and quenched inside water for few seconds. The ampoule is then annealed at a temperature slightly below the glass transition temperature ($T_g$) around 165°C. In the next step, the chalcogenide glass rod is retrieved by breaking and removing the ampoule and put in a two-chamber distillation tube in order to purify it through several distillation processes. In the first distillation step, distillation tube is placed inside a localized heater at 600°C and the glass rod is distilled under dynamic vacuum to remove impurity gas like HCl and CCl$_4$ residue [40]. In the second step, static distillation is performed at 640°C and carbon, silica and other refractory oxides are removed from the glass. At the end of second distillation step, the glass is again homogenously heated inside a furnace for 10 hours at 850°C, then the ampoule is cooled down to 500°C, quenched and finally annealed below T$_g$ to get the final glass. The attenuation spectrum of bulk As$_{38}$Se$_{62}$ glass measured with a Bruker Tensor 37 FTIR spectrometer is shown in Fig. 1. This glass offers light transmission from 2-9 µm range with a minimum attenuation of 0.40 dB/m at 7.42 µm and an absorption peak at 4.56 µm due to the presence of Se-H chemical bonds.

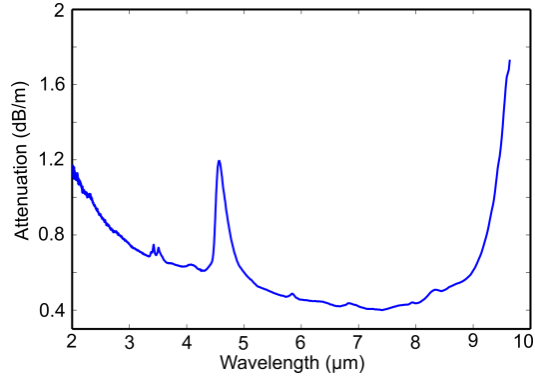

Fig. 1. Attenuation spectrum of bulk As$_{38}$Se$_{62}$ glass.

*2.2 Preform fabrication*

The PM microstructure preform containing 3-rings of air holes shown in Fig. 2(a) is prepared using the molding method [43, 37]. In this method, the above-synthesized highly purified As$_{38}$Se$_{62}$ glass is heated and flowed into a silica mold, which has the negative shape of the final preform. The silica mold is made of silica-glass capillaries with a thickness of 30 µm. These capillaries are organized in a hexagonal periodical pattern, using slices of a silica microstructured preform to keep them in place. The hole diameter is 460 µm (d$_{hole}$) and the pitch is 1350 µm (Λ). In order to make a PM preform, the geometry of slices is modified by placing two larger holes with a diameter of 650 µm in diametrically-opposed positions in the first ring, while maintaining the d$_h$/Λ ratio. Both the As$_{38}$Se$_{62}$ glass rod and the silica mold are placed in a silica ampoule. The ampoule is then heated in a rocking furnace at 600 °C for one hour so the chalcogenide glass can flow down in to the mold having a glass viscosity about $10^{-4}$ Pa.s. After that, air quenching is performed on the full mold for four minutes before annealing it at 165°C for one hour. The silica tube around the preform is then removed with a diamond tool. At the final step, the chalcogenide preform is obtained by removing the embedded silica capillaries with a hydrofluoric acid treatment. The final diameter of the As$_{38}$Se$_{62}$ glass preform is around 16 mm having a length of 80 mm.

*2.3 Fiber fabrication*

For the drawing process, the As$_{38}$Se$_{62}$ preform is installed in the drawing tower inside a silica glass enclosure, through which helium gas is flowed to remove any remaining moisture from inside it. An annular electrical furnace is placed around the preform in such a way that it helps to heat the lower portion of the preform. Both the preform chamber and the furnace chamber were filled with helium gas. The temperature of the furnace is increased until a drop of glass fall down from the preform due to the gravitational force (around 340°C). Then the fiber accompanying the drop is attached to a rotating drum to continue the drawing while the preform is being fed gradually inside the annular furnace in the same time. During the drawing process, the desired air-hole diameter is obtained by controlling the helium gas pressure inside enclosure. The diameter of the fiber is further controlled by the pulling speed of the rotating drum and the feeding speed of the preform into the furnace.

Figures 2(b) and (c) show the scanning electron microscope (SEM) images of the PM-PCF drawn from the preform. As can be seen, it consists of 36 circular air holes in 3 rings with 2 larger air holes adjacent to the core that provides strong birefringence and polarization maintaining guiding. The distance between the closest points of the two large air holes is 8.11 µm and the outer diameter of the PCF is 125 µm. The diameter of the large air holes ($d_{LA}$) and small air holes ($d_{SA}$) are 5.8 µm and 3.35 µm, respectively. The distance between two small air holes, also known as the pitch is 7.025 µm, which gives a $d_{SA}/\Lambda$ ratio of 0.477, which makes the PCF with uniform holes endlessly single-moded in the mid-IR [44].

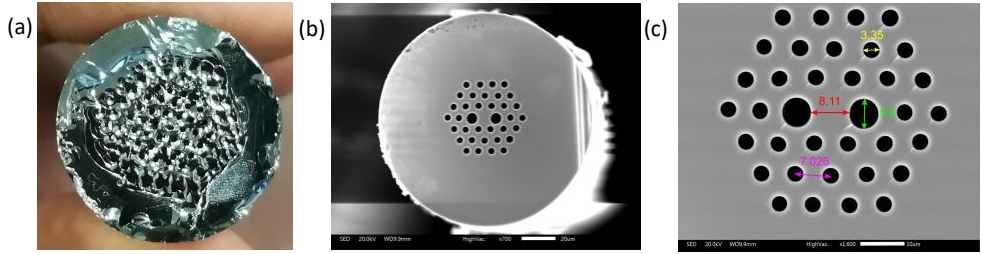

Fig. 2. Cross-section images of $As_{38}Se_{62}$ preform and solid-core microstructured fiber. (a) Preform of $As_{38}Se_{62}$ glass. (b) SEM image of $As_{38}Se_{62}$ PM-PCF with an outer diameter of 125 µm. (c) Expanded view of the microstructure region showing core diameter ($d_{core}$ = 8.11 µm), small air holes diameter ($d_{SA}$ = 3.35 µm) and pitch ($\Lambda$ = 7.025 µm).

## 2.4 Taper fabrication

$As_{38}Se_{62}$ PM-PCF tapers were directly made on the drawing tower using the following method. One end of a previously drawn PM fiber (as described in Sec. 2.3) with a length of 80 cm is fixed at the top of the drawing tower. In this case, the outer fibre diameter was 162 µm. Finally, the taper is obtained by drawing a second time the fiber.

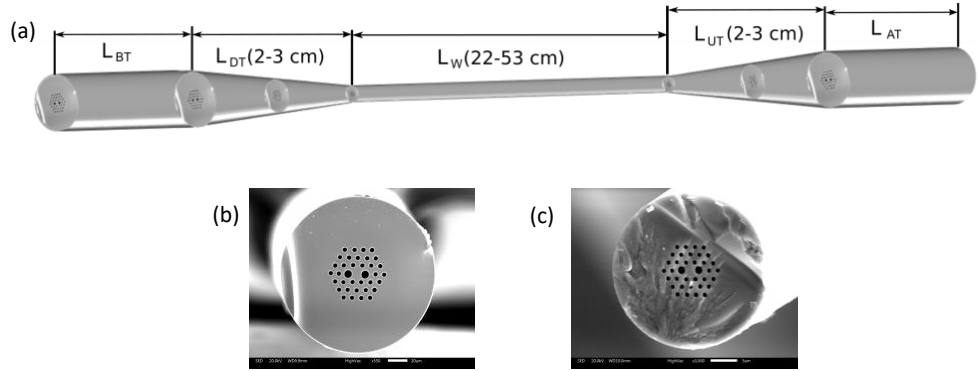

Fig. 3 (a). Schematic of longitudinal sections of $As_{38}Se_{62}$ PM-PCF tapers: Length before taper ($L_{BT}$), length of down-taper section ($L_{DT}$), Length of uniform waist section ($L_W$), Length of up-taper section ($L_{UT}$), and length after taper ($L_{AT}$). (b) & (c) Cross-section images of normal section and taper-waist section of $As_{38}Se_{62}$ PM-PCF with respective outer diameter of 161.9 µm and 28.8 µm and respective core diameter of 9.97 µm and 1.77 µm.

Table 1. Structural parameters of a PM-PCF taper

| Initial fiber | | | | | Taper waist | | | | |
|---|---|---|---|---|---|---|---|---|---|
| $d_{core}$ [µm] | $d_{out}$ [µm] | $d_{SA}$ [µm] | $\Lambda$ [µm] | $d_{SA}/\Lambda$ | $d_{core}$ [µm] | $d_{out}$ [µm] | $d_{SA}$ [nm] | $\Lambda$ [µm] | $d_{SA}/\Lambda$ |
| 9.97 | 161.9 | 4.25 | 8.51 | 0.499 | 1.77 | 28.8 | 808.03 | 1.69 | 0.478 |

Since the initial PM fiber has an outer diameter of 162 µm and core diameter of 10 µm, the output diameter of the waist should be 28.74 µm in order to have an approximate core diameter of 1.77 µm in the taper waist region. For this purpose, the pulling speed and the feeding speed are fixed at 0.05 m/min and 1.5 mm/min, respectively. The obtained PM-PCF tapers have a down-taper section length of 2-3 cm, waist section length of 22-53 cm, and up-taper section length of 2-3 cm. Figures 3(a) and (b) represent the SEM image of the normal section and tapered section of $As_{38}Se_{62}$ PM-PCF taper. The geometrical parameters of this taper are given in Table. 1, where $d_{core}$, $d_{out}$, $d_{SA}$, and P represent waist core diameter, outer fiber diameter, small air hole diameter and pitch, respectively. Proper maintenance of $d_{Sa}/\Lambda$ ratio during the taper fabrication provides closer values of 0.499 and 0.478 for the normal section and the taper section, respectively.

## 3. Computation of dispersion and birefringence

Dispersion and birefringence properties of the PM-PCF and the PM taper are computed from the cross-section images shown in Figs. 2(c) and 3(c). We calculated the effective refractive index of the fundamental mode using COMSOL software based on a full vector finite-element method. The PM-PCF fibers are assumed to be single-mode ($HE_{11}$) as the second-order and higher-order modes have much higher confinement loss (18 dB/m at 3 µm, 56 dB/m at 4 µm, and 176 dB/m at 5µm) than $HE_{11}$ mode over the wavelength range under investigation. The refractive index of the air holes is set to 1 and the refractive index of the glass was calculated using a standard Sellmeier equation with the following form [45]

$$n^2(\lambda) = A_0 + \frac{A_1 \lambda^2}{\lambda^2 - a_1^2} + \frac{A_2 \lambda^2}{\lambda^2 - a_2^2}, \qquad (1)$$

where, $A_0$, $A_1$, and $A_2$ are dimensionless coefficients and $a_1$ and $a_2$ are the material resonant wavelengths. For $As_{38}Se_{62}$ glass, the values of the coefficients used in the simulation are: $A_0 = 3.7464$, $A_1 = 3.9057$, $A_2 = 0.9466$, $a_1 = 0.4073$ µm, and $a_2 = 40.082$ µm [37]. The effective refractive index ($n_{eff}$) is then used to calculate the GVD, defined as $D = -(\lambda/c)(d^2 n_{eff}/d\lambda^2)$ and the phase birefringence, defined as $B(\lambda) = (n_{eff}^y(\lambda) - n_{eff}^x(\lambda))$, where $n_{eff}^x$ and $n_{eff}^y$ are the effective refractive indices of x-polarized $HE_{11}$ mode (slow axis) and of y-polarized $HE_{11}$ mode (fast axis). The computed GVD and phase birefringence are plotted in Fig. 4(a) for the PM-PCF with a core diameter of 8.11 µm and in Fig. 4(b) for a PM-PCF taper with a waist diameter of 1.77 µm. The PM-PCF has a zero-dispersion wavelength (ZDW) around 4.55 µm for the fast axis and, at this wavelength, a large birefringence of $6.10^{-4}$ such that strong polarization maintaining properties are expected. This is in excellent agreement with the experimental measurement recently reported in Ref. [43]. By tapering the PM-PCF core from 8.11 µm to 1.77 µm, we are able to bring down the ZDW from 4.5 µm to 2.507 µm along with a second ZDW at 3.51 µm. The PM-PCF taper also gives higher phase birefringence of $7.97 \times 10^{-3}$ and $1.79 \times 10^{-2}$ at two respective ZDWs.

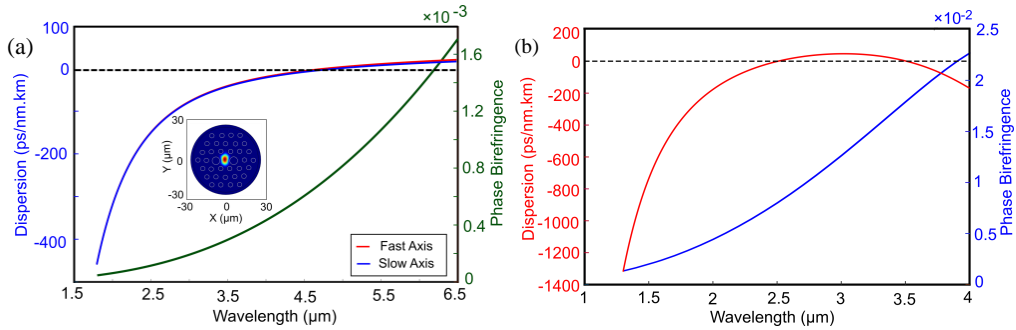

Fig. 4. Dispersion and birefringence characteristics of the PM-PCF simulated for the fundamental mode from SEM image. (a) PM-PCF with core diameter of 8.11 µm. Blue and red curve: group velocity dispersion of PM-PCF sample for different polarization axis. Inset: optical power density of the fundamental mode for the fast axis at a wavelength of 2 µm inside the core of the fiber. Green curve: phase birefringence of PM-PCF. (b) PM-PCF taper with a waist core diameter 1.77 µm. Red curve: group velocity dispersion of taper-waist section. Blue curve: phase birefringence of the taper-waist section.

## 4. Optical transmission and attenuation

Optical attenuation of the drawn PM-PCF with a core diameter of 8.11 µm was measured with a Bruker Tensor 37 FTIR spectrometer equipped with a liquid nitrogen cooled mercury-cadmium-tellurite (MCT) detector having a sensitivity from 1.2-14 µm, as shown in Fig. 5(a). The attenuation of the fiber was precisely measured using the standard cut-back technique and guidance of cladding mode is eliminated by applying a layer of graphite on the surface of the fiber during the measurement. Figure 5(a) shows light transmission in the PM-PCF between 2 µm and 10 µm with a minimum attenuation of 0.25 dB/m at 6.64 µm. Two absorption peaks can be observed in the attenuation spectrum of the fiber including a peak at 4.5 µm due to the presence of Se-H bonds (as in the bulk sample, See Fig. 1) and another peak at 6.3 µm due to the presence of $H_2O$ molecules.

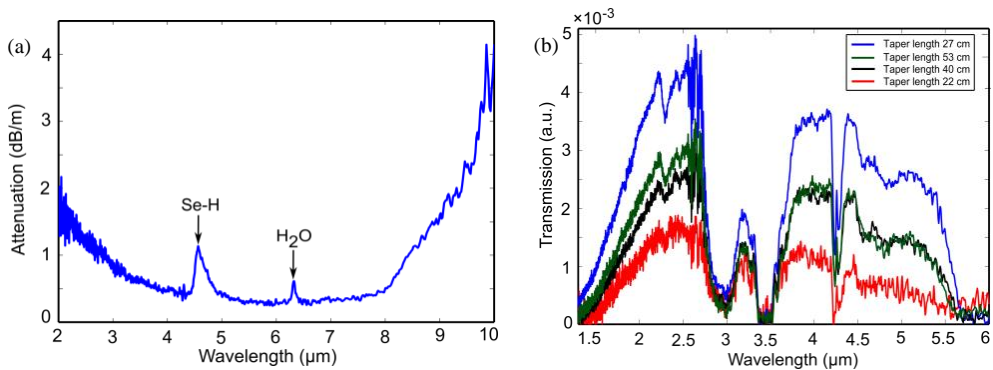

Fig. 5 (a). Attenuation spectrum of $As_{38}Se_{62}$ PM-PCF with cut-back length of 3.35 m. (b) Transmission spectra of $As_{38}Se_{62}$ PM taper fibers with different taper waist length having a waist diameter 1.77 µm.

The power transmission of a tapered PCF with taper waist core diameter of 1.77 µm and taper waist length of 22 cm was measured to be 13 % at 1.55 µm using a 1mW continuous wave (CW) laser and a Thorlabs power meter. Then, the transmission spectra of all four tapers are measured with the same FTIR spectrometer as previous, which is shown in Fig. 5(b). The tapers transmit light from 1.5-5.5 µm and there is no significant change in transmission window with taper length. As can be seen, the transmission bandwidth is significantly reduced compared to the untapered PM-PCF and there are three transmission dips clearly visible because of the presence of three new absorption bands at 2.9 µm due to O-H bonds, at 3.5 µm due to surface contamination by C-H Bonds of non-cyclic organic compounds, and at 4.25 due to $CO_2$ molecules. These are mainly due to the contamination of the taper waist, non-adiabatic taper transitions during taper drawing process, and high confinement loss for non-uniform and small taper waist. This unfortunately prevents the use of the PM tapers for mid-IR SC generation beyond 5.5 µm and we will therefore in the following sections focus on the straight PM-PCFs only.

## 5. Experimental setup for supercontinuum generation

Figure 6 outlines the experimental setup for mid-IR pumping of the chalcogenide-based PM-PCF and measuring the output SC infrared light [5]. The experiment was performed at the Technical University of Denmark. A 10 mm periodically-poled fan-out $MgO:LiNbO_3$ crystal (MgO:PPLN) was used to generate the mid-IR pump through a single-pass parametric generation process. The nonlinear crystal was pumped by a continuous-wave (CW) seed laser (tunable from 1350 to 1450 nm) along with a 1.04 µm mode-locked Yb: KYW solid-state laser (having a full width half maximum pulse duration of 250 fs at 21 MHz repetition rate) to stimulate quasi phase-matched parametric anti-Stokes generation. The phase-matching relation was achieved by selecting the appropriate poling period of the fan-out structure using a linear translation stage, and the crystal was kept in an oven at a constant temperature of 150°C to avoid photorefraction. The combination of tunable seed and tunable poling period of the crystal provides a tunable MIR output between 3.7 µm and 4.53 µm whose spectrum corresponds to a 252 fs transform limited pulse train at 4 µm (see [46] for the first use of the laser). A reflective long pass filter (LPF) was used after the crystal system in order to eliminate any residual pump and other radiation below 3.5 µm. The mid-IR output beam was then collimated by an achromatic air-space lens doublet, which is optimized for a central wavelength of 4 µm and anti-reflection (AR) coated for the 3-5 µm range. The pump beam was then injected into the PM-PCF through an AR coated ZnSe aspheric lens with a focal length of 12 mm and the coupling power was controlled by rotating a wire grid polarizer relative to the linear polarization of the pump. The generated SC light was measured using an FTIR from 3-10 µm and a monochromator-based spectrometer from 1-5 µm. The output power was measured with a thermal power meter.

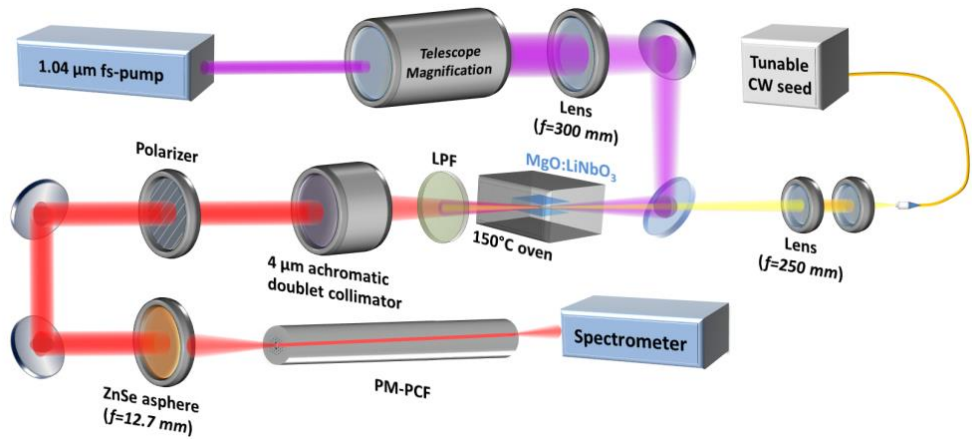

Fig. 6. Scheme of the experimental setup for mid-IR pump laser and supercontinuum infrared light generation. LPF, long-pass filter; PM-PCF, polarization maintaining photonic crystal fiber.

## 6. Results and discussions

Figure 7(a) shows the generated SC spectra with increasing pump power in a 25 cm long PM-PCF with the pump tuned to 4 µm central wavelength. For a pump power of 135 mW, we obtained a SC spectrum spanning from 3.1-6.02 µm (at the -30 dB level) with an average output power of 11 mW.

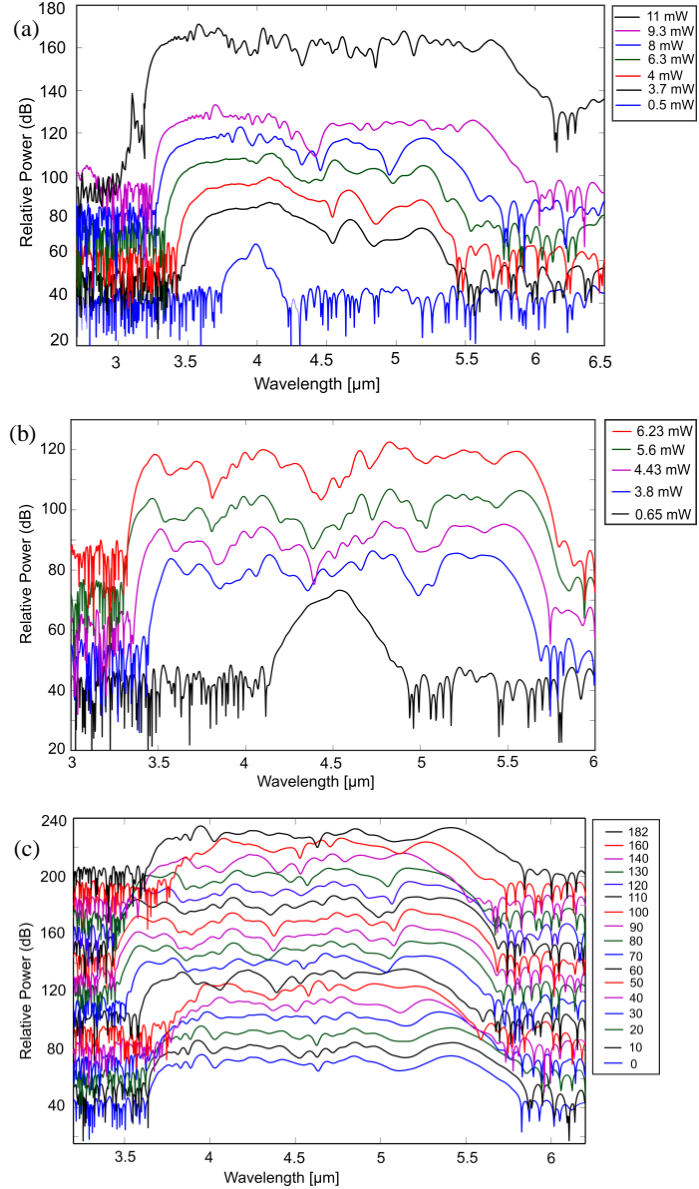

Fig.7. Supercontinuum spectra generated in a 25 cm long PM-PCF sample with 252 fs pulses as a function of mean output power. (a) SC spectra generated with a 4 µm pump. (b) SC spectra generated with a 4.53 µm pump. (c) Dependency of SC spectra with incident angle of polarization from 0° to 182° for pumping at 4.53 µm.

Then we tuned the pump wavelength close to the ZDW at 4.53 µm and the obtained SC spectra with different output power are shown in Fig. 7(b). An available maximum power of 98 mW provided a spectral broadening from 3.33-5.78 µm (at the -30 dB level) with an average output power of 6.23 mW. Therefore, the spectrum at 4.53 µm has a lower bandwidth than the spectrum at 4 µm due to available lower input power from the laser in spite of pumping closer to the ZDW. Furthermore, with a constant input power of 62 mW at 4.53 µm, we were able to measure the polarization dependency of the SC spectra by rotating the input fiber, which amounts to turning the input pump polarization with respect to the principal axes of the PM fiber. Using the measured output power, Fresnel loss from both front and end surface of the fiber, and propagation loss inside the fiber, we estimated the coupling efficiency to be 18% and 13.62 %, the coupling peak power as 4.26 kW and 1.54 kW at 4 µm pumping and 4.53 µm pumping, respectively. Figure 7 (c) shows the recorded SC spectra when tuning the fiber angle from 0° till 182° with 10° steps and with constant output power of 3.8 mW. As it can be seen, the input pump polarization strongly affects the SC generation. The SC bandwidth reduces from 2260 nm down to 1800 nm (at the -20 dB level) with an angular periodicity of about 90°.

## 7. Numerical simulations: The vector model

Detailed modelling of the polarization and modal properties of mid-IR SC generation in multi-mode step-index non-PM chalcogenide fibers was studied in [47], which demonstrated how important it is to control the polarization and the number of modes participating in the SC generation process. Here the uniform PM-PCF is single-moded, which means that we can restrict ourselves to consider only the fundamental mode with both polarizations. Nonlinear pulse propagation and SC generation in the highly-birefringent PM-PCF was therefore modelled using a two-polarization code based on two coupled generalized nonlinear Schrödinger equations (CGNLSE) [48, 49]. These equations can be written in following reduced form

$$\frac{\partial A_{x,y}}{\partial z} + \frac{\alpha}{2} A_{x,y} + \beta_{1x,y} \frac{\partial A_{x,y}}{\partial T} - \sum_{k \geq 2} \frac{i^{k+1}}{k!} \beta_k^{x,y} \frac{\partial^k A_{x,y}}{\partial T^k}$$
$$= i\gamma \left(1 + i\tau_{shock} \frac{\partial}{\partial T}\right)$$
$$\times \left(A_{x,y}(z,T) \int_{-\infty}^{+\infty} R(T') |A_{x,y}(z, T-T')|^2 dT'\right), \quad (2)$$

where $A_x$ and $A_y$ are the slow and fast axis field amplitudes, $\alpha$ is the wavelength-dependent propagation loss taken from Fig. 5(a), $\beta_k^{x,y}$ is the k$^{th}$ Taylor expansion of the propagation constants for the two cross-polarized fields. The second-order $\beta_2^{x,y}$ is shown in Fig 4(a) as the $D$ parameter. Nonlinear coefficient $\gamma$ was calculated from effective mode area of PM-PCF and nonlinear index of the $As_{38}Se_{62}$ glass using $\gamma = \frac{2\pi n_2}{\lambda A_{eff}(\lambda)}$. This gives $\gamma$ = 358.6 W$^{-1}$km$^{-1}$ and 301 W$^{-1}$km$^{-1}$ at the two pumping wavelengths, 4 µm and 4.53 µm, respectively [50]. The time derivative term in RHS of Eq. (2) models the effects such as self-steepening and optical shock formation with $\tau_{shock} = \frac{1}{\omega_0}$. The term $R(T) = (1 - f_R)\delta(t) + f_R h_R(t)$ represents the response function which includes both instantaneous electronic Kerr $\delta(t)$ and delayed Raman $h_R(t)$ contribution. $f_R$ is the Raman fractional contribution to the Kerr effect and we find $f_R = 0.1$ using the parameter reported in Ref. [51] for chalcogenide glass. The Raman response delayed function $h_R(t)$ can be expressed in the time as follows:

(3)

$$h_R(t) = \frac{\tau_1^2 + \tau_2^2}{\tau_1^2 \tau_2^2} \exp\left(\frac{-T}{\tau_2}\right) \sin\left(\frac{T}{\tau_1}\right),$$

where $\tau_1$ and $\tau_2$ are 23.14 fs and 157 fs, respectively, which was fitted for a Raman gain peak of 230 cm$^{-1}$ (6.9 THz) for $As_{38}Se_{62}$ fiber whose Raman gain peak was measured in Ref. [52]. The same Raman function was used for both axes because the perpendicular Raman gain is almost negligible [53]. Note that in numerical code, we have also considered the degenerate cross-phase modulation (DXPM) between the two cross-polarized fields with the factor 2/3 [53]. For the PM-PCF sample with pump at 4 µm and 4.53 µm, we used a fiber length of 25 cm and pump pulse duration of 252 fs for both cases. The coupling peak power at the input of the fiber was considered to be 1.33 kW and 0.40 kW for the pump at 4 µm and 4.53 µm, respectively, in order to have the best correspondence with experimental results.

Figures 8(b) and (d) shows the evolution of numerically computed SC spectra along fiber length for a pump wave polarized along one of the principal axis at 4 µm and 4.53 µm, respectively. The comparisons between simulated and experimental SC spectra for respective pump wavelength of 4 µm and 4.53 µm are shown in Figs. 8(a) and (c). The simulated and experimental SC spectra are in qualitative agreement in terms of bandwidth, but in terms of structure of the spectrum, they are quite different.

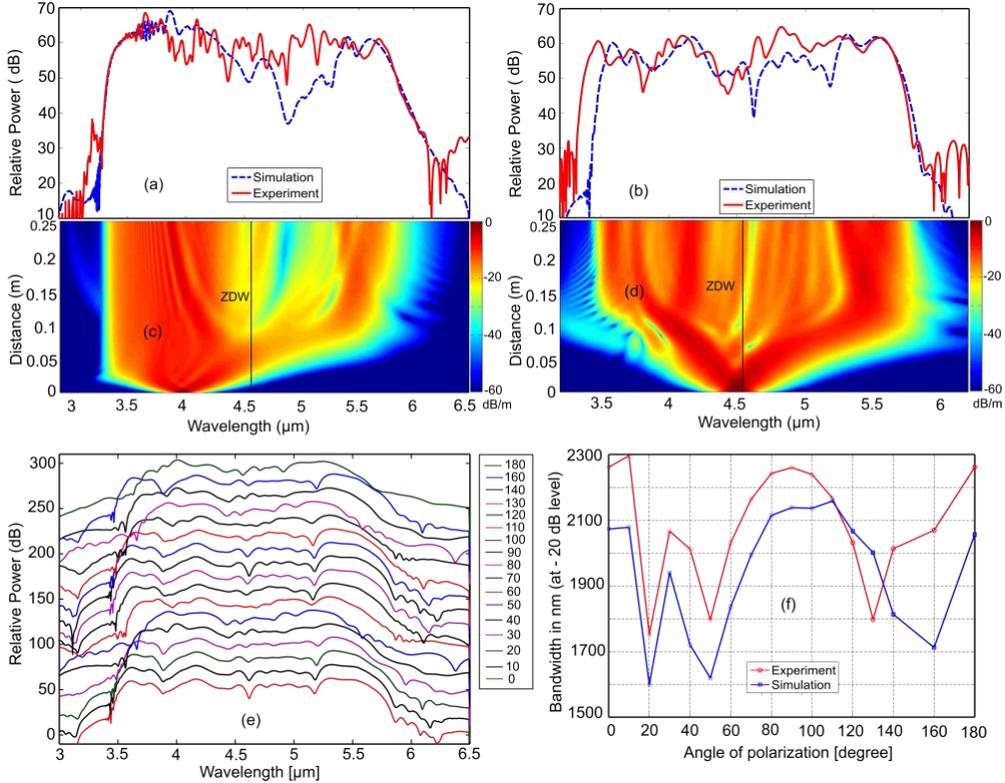

Fig. 8. (a) & (b) Comparison between numerically and experimentally generated SC spectra in the PM-PCF for pumping at 4 µm (a) and 4.53 µm (b). (c) & (d) Numerically generated evolution of SC spectra along the fiber length at the respective pumping wavelength. ZDW, zero-dispersion wavelength. (e) Polarization dependency of numerically generated SC spectra at 4.53 µm pumping. (f) Comparison of bandwidth (at -20 dB level) between experimental and simulated SC spectra at different polarization angle for 4.53 µm pumping.

In Figs. 8(c) and (d), we plotted the SC dynamics along the propagation distance and we can clearly observe multiple soliton generation and ejection towards the infrared up to 6 µm due the Raman-self frequency shift. Simultaneously there are dispersive waves that are emitted from solitons in the short wavelength range down to 3 µm.

Figure 8(e) shows a series of computed SC spectra as a function of input polarization angle $\theta$, from 0 to 180°. This leads to SC bandwidth reduction from 2160 nm down to around 1600 nm with an angular periodicity of around 90°, as shown in Fig. 8(f) as a blue curve. The agreement with experimental measurements (red curve) is quite satisfactory. This can simply be explained by the fact that when tuning the polarization angle from one principal axis to the other, the degenerate cross-phase modulation between the $A_x$ and $A_y$ vector fields is lower than the self-phase modulation (SPM) itself by a factor 2/3, giving rise to less SC broadening when pumping at 45°.

## 8. Conclusion

In summary, polarization-preserving supercontinuum generation in the mid-infrared from 3.1 to 6.02 µm and from 3.33 to 5.78 µm have been demonstrated by pumping a highly-birefringent photonic crystal fiber made of chalcogenide glass with femtosecond pulses at 4 µm and 4.53 µm, respectively. It has been further shown that tuning the input polarization angle with respect to the fiber axes affects the supercontinuum bandwidth and this was checked against numerical simulations using vectorial two-coupled nonlinear Schrödinger equations. The fiber was drawn using the casting method from $As_{38}Se_{62}$ glass which features a transmission window from 2 to 10 µm and a high nonlinear index of $1.13 \times 10^{-17}$ m$^2$/W. It has a zero-dispersion wavelength around 4.5 µm and, at this wavelength, a large birefringence of $6.10^{-4}$ enabling strong polarization maintaining properties. Long tapered optical fibers have also been designed and fabricated from the same fiber with waist diameter down to a 1.77 µm and with ultra-high birefringence up to $10^{-2}$. Finally, this work constitutes an important step towards stable and linearly-polarized supercontinuum generation for mid-IR applications.

## 9. Funding

European Union's Horizon 2020 Research and Innovation Programme under grant agreement n°722380; Agence Nationale de la Recherche (ANR) (ANR-15-IDEX-0003, ANR-17-EURE-0002), OB and CRP acknowledge support by the European Union's Horizon 2020 research and innovation program under grant agreement No. 732968 project FLAIR, and from Innovation Fund Denmark, project ShapeOCT (J. No. 4107-00011A).